# CHAOS AND ELLIPTICAL GALAXIES

D. MERRITT

*Department of Physics and Astronomy, Rutgers University*

**Abstract.** Recent results on chaos in triaxial galaxy models are reviewed. Central mass concentrations like those observed in early-type galaxies – either stellar cusps, or massive black holes – render most of the box orbits in a triaxial potential stochastic. Typical Liapunov times are 3-5 crossing times, and ensembles of stochastic orbits undergo mixing on time scales that are roughly an order of magnitude longer. The replacement of the regular orbits by stochastic orbits reduces the freedom to construct self-consistent equilibria, and strong triaxiality can be ruled out for galaxies with sufficiently high central mass concentrations.

## 1. Introduction

Stellar dynamics was one of the first fields where chaos was understood to be important, but the relevance of chaos to galaxies has never been completely clear. Three arguments have commonly been cited to show that chaos is unlikely to be important in determining the structure of real galaxies. First, it is pointed out that many reasonable potentials contain only modest numbers of stochastic orbits. This is always true for the potentials of rotationally symmetric models, and there is even a class of non-axisymmetric models for which the motion is completely regular, including the famous "Perfect Ellipsoid" (Kuzmin 1973; de Zeeuw & Lynden-Bell 1985). Second, it is noted that stochastic orbits often behave much like regular orbits over astronomically interesting time scales. Therefore (it is argued) one need not make a sharp distinction between regular and stochastic orbits when constructing an equilibrium model. Third, following the successful construction by Schwarzschild (1979, 1982) and others of self-consistent triaxial equilibria, it has generally been assumed that the regular orbits – which are confined to narrow parts of phase space and thus have definite "shapes" – are the only useful building blocks for real galaxies.

Recent theoretical work, combined with an improved understanding of the central structure of early-type galaxies, have exposed weaknesses in each of these arguments. The phase space of a triaxial model that looks similar to real elliptical galaxies is typically strongly chaotic; the regular box orbits (the "backbone" of elliptical galaxies) usually do not exist. Furthermore the stochastic orbits in such models often behave chaotically on time scales that are short compared to the age of the universe; typical Liapunov times are just 3-5 orbital periods, and characteristic mixing time scales are roughly ten times longer. A few attempts have now been made to build self-consistent triaxial models with realistic density profiles, and one finds that a large fraction of the mass must typically be placed on the stochastic orbits. In models with strong central mass concentrations, the generally spherical shape of the stochastic orbits, combined with their dominance of phase space, can preclude a self-consistent equilibrium.

Here I review this work and present some speculations on its relevance for the structure and evolution of elliptical galaxies.

## 2. Imperfect Ellipsoids

Kuzmin (1973) showed that there is a unique, ellipsoidally-stratified mass model for which the corresponding potential has three global integrals of the motion, quadratic in the velocities. The density of Kuzmin's model is given by

$$\rho = \rho(m) = \frac{\rho_0}{(1+m^2)^2}, \qquad m^2 = \frac{x^2}{a^2} + \frac{y^2}{b^2} + \frac{z^2}{c^2}, \qquad (1)$$

with $a \geq b \geq c$ the axis lengths that define the ellipsoidal figure. The oblate case, $a = b$, was discovered by Kuzmin already in 1956 in his classic study of separable models of the Galaxy; the fully triaxial case was rediscovered by de Zeeuw & Lynden-Bell in 1985, who christened Kuzmin's model the "Perfect Ellipsoid." Kuzmin (1973) and de Zeeuw (1985) showed that all the orbits in the Perfect Ellipsoid potential fall into one of only four families. Three of these families are "tube" orbits that respect an integral analogous to the angular momentum about the long $(x)$ or short $(z)$ axis; tube orbits avoid the center, as do almost all of the orbits in an arbitrary axisymmetric potential. The fourth family of orbits, the "boxes," are unique to triaxial potentials. Box orbits have filled centers and touch the equipotential surface at eight points, one in each octant. Self-consistent triaxial models (Schwarzschild 1979; Statler 1987) tend to weight the box orbits heavily, since these orbits have mass distributions that mimic that of the underlying mass model.

Kuzmin's density law was arrived at via mathematical manipulations and it is hardly surprising that the Perfect Ellipsoid bears little resemblance to the distribution of light or mass in real elliptical galaxies. The discrepancy is particularly great near the center, where Kuzmin's law predicts a large, constant-density core. The luminosity densities in real elliptical galaxies are always observed to rise monotonically at small radii (Ferrarese *et al.* 1994; Møller *et al.* 1995; Lauer *et al.* 1995). The steepest cusps are seen in low-luminosity galaxies like M32, which has $\rho \propto r^{-1.63}$ near the center (Gebhardt 1995). More luminous ellipticals like M87 have cusps with power-law indices flatter than $-1$ (Merritt & Fridman 1996b). The cusps in these luminous galaxies are sufficiently shallow that the power-law nature of the profile is obscured when the cusp is observed through the outer layers of the galaxy; as a result, such galaxies are sometimes still described as having "cores" (e.g. Kormendy *et al.* 1995) even though their central densities diverge as power laws.

Although the Perfect Ellipsoid does not describe real galaxies very well, its complete integrability makes it useful as a starting point for constructing more realistic models. Consider the mass model

$$\rho(m) = \frac{\rho_0}{(m_0^2 + m^2)(1 + m^2)}, \qquad (2)$$

which might be called the "Imperfect Ellipsoid." For $m_0 = 1$, Eq. (2) reduces to Kuzmin's law, while for $m_0 = 0$ the Imperfect Ellipsoid has a $\rho \propto r^{-2}$ central density cusp like those observed in some elliptical galaxies. Thus varying $m_0$ from 1 to 0 takes one from a fully integrable but nonphysical model, to a strongly nonintegrable but realistic model.

Every box orbit in the Perfect Ellipsoid will eventually pass arbitrarily close to the center, and a small enough value of $m_0$ might be expected to destroy the integrability of at least some of these orbits. The dissolution of the box orbits can be tracked by computing their Liapunov characteristic numbers, defined in the usual way as finite-time averages of the rate of exponential divergence of intially close trajectories. Of the three positive Liapunov numbers, one is always close to zero; the other two Liapunov numbers $\sigma_1$ and $\sigma_2$ should tend to zero only if the orbit is regular. Fig. 1 shows the distributions of $\sigma_1$ and $\sigma_2$ for sets of 192 orbits that were dropped with zero velocity from an equipotential surface in the potential corresponding to Eq. (2), with $c/a = 0.5$, $b/a = 0.79$ and various values of $m_0$. The orbits have energies equal to that of the long-axis orbit that just touches the ellipsoidal shell dividing the model into two equal-mass parts. The Liapunov numbers were computed over $10^2$, $10^3$ and $10^4$ dynamical times $T_d$, defined as the full period of the long-axis orbit.

A large fraction of the orbits with box-like initial conditions are stochastic, and this fraction is not tremendously dependent on $m_0$. Even for $m_0 = 0.1$ — which is much too large to describe real elliptical galaxies — more than half of the equipotential surface generates stochastic orbits, and this fraction increases to $\sim 150/192$ when $m_0$ has the physically more realistic values of $10^{-2}$ or $10^{-3}$.

Fig. 1 shows that the distribution of Liapunov numbers for the stochastic orbits always evolves toward a single narrow peak – presumably because every stochastic orbit at a given energy moves in the same stochastic sea. One can therefore identify, after sufficiently long integrations, the unique numbers $\sigma_1(E)$ and $\sigma_2(E)$ that characterize stochastic motion at any given energy in these models. Expressed in units of the inverse dynamical time $T_d$, Fig. 1 shows that $\sigma_1 \approx 0.2 - 0.3$ and $\sigma_2 \approx 0.04 - 0.1$. Thus, divergence between nearby stochastic orbits takes place on time scales that are only $3 - 5$ times longer than the dynamical time.

## 3. Central Singularities

These results demonstrate that chaos can be important in triaxial models without singular densities or divergent forces. However there is increasingly strong evidence that early-type galaxies *do* contain central singularities, in the form of massive dark objects, possibly black holes (Ford *et al.* 1994; Miyoshi *et al.* 1995). Fig. 2 shows how the distribution of Liapunov numbers in the Imperfect Ellipsoid with $m_0 = 0.1$ is changed by the addition of a central point mass containing $10^{-2}$, $3 \times 10^{-3}$ or $10^{-3}$ of the total galaxy mass. (The ratio of black hole mass to luminous galaxy mass is thought to be about $5 \times 10^{-3}$ for M87 and $2.5 \times 10^{-3}$ for M32.) Even the smallest of these "black holes" induces the majority of the box-like orbits to behave chaotically, with $\sigma_1 T_d \approx 0.15$ or greater.

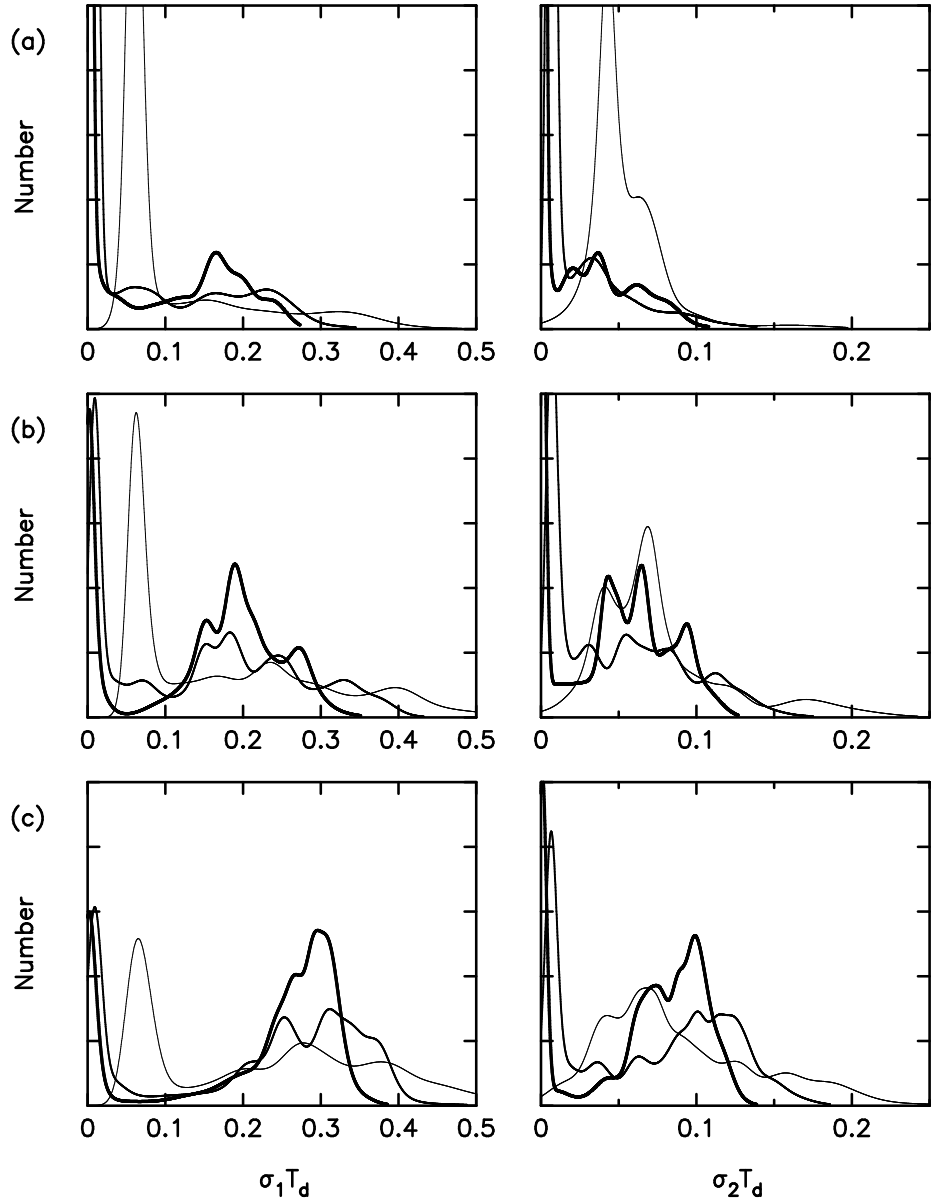

Fig. 1. Histograms of Liapunov numbers for isoenergetic ensembles of box-like orbits in the triaxial potential corresponding to Eq. 1, with $m_0 = 10^{-1}$ (a), $m_0 = 10^{-2}$ (b) and $m_0 = 10^{-3}$ (c). $c/a = 0.5$ and $b/a = 0.79$. Orbits were integrated for $10^2$ (thin lines), $10^3$ (medium lines) and $10^4$ (solid lines) dynamical times.

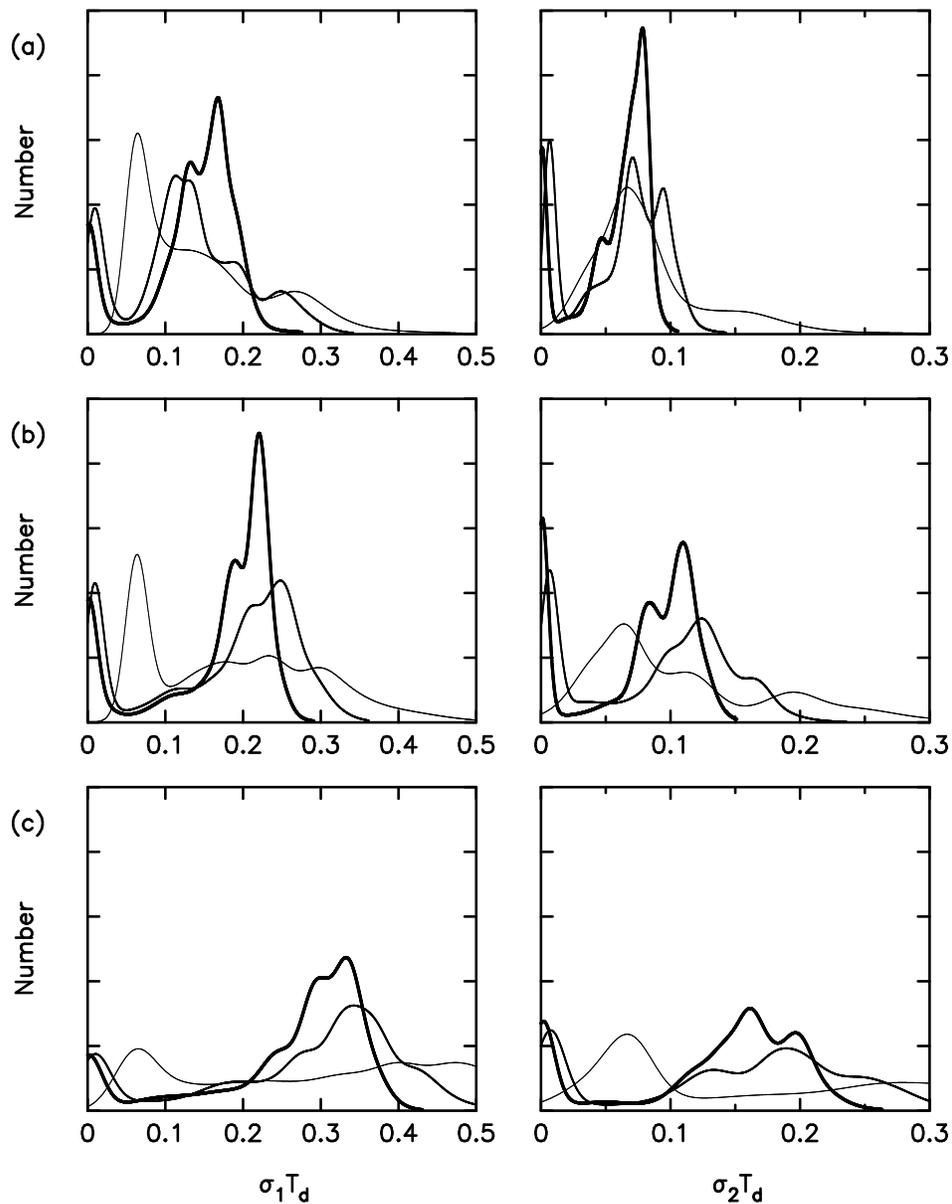

Fig. 2. Like Fig. 1, for orbits in the potential of Eq. 1 with $m_0 = 10^{-1}$. A central point mass containing 0.1% (a), 0.3% (b) and 1% (c) of the galaxy mass has been included.

Singular forces also occur in galaxies where the stellar density increases more rapidly than $1/r$ toward the center, as is the case in low-luminosity ellipticals (Kormendy *et al.* 1995). Schwarzschild (1993) and Merritt & Fridman (1996a) have investigated the orbital structure of triaxial potentials with $\rho \propto r^{-1}$ and $r^{-2}$ density cusps and find that the box-like phase space is largely chaotic. The long-axis orbit, which generates the box orbits in integrable triaxial potentials, is unstable at most energies when the central density increases more rapidly than $r^{-0.5}$ (Merritt & Fridman 1996b). Much of the stochasticity is driven by instabilities out of the principal planes and was missed in earlier studies that restricted the motion to two dimensions.

It is possible that the majority of elliptical galaxies contain either a central black hole, a steep central cusp, or both. If so, then chaos and triaxiality are strongly linked.

## 4. Mixing

An important quantity is the time required for an initially non-random distribution of stars in stochastic phase space to "mix" into a time-invariant state. This process is essentially irreversible, and if it occurs on a time scale that is much shorter than the age of a galaxy, we would expect the stochastic parts of phase space to have reached a nearly steady state by now. On the other hand, if the mixing time exceeds a galaxy lifetime, then triaxial galaxies might still be slowly evolving as the stochastic orbits continue to mix.

Astronomers often talk loosely about "phase mixing" of regular orbits, but dynamicists discuss mixing only in the context of chaos. In fact one can define a hierarchy of dynamical systems in terms of their degree of irreversibility or randomness (e.g. Zaslavsky 1985). The lowest rung on the ladder is occupied by "ergodic" systems; an ergodic trajectory has the property that the average time spent in any phase-space volume is proportional to that volume. Thus an ergodic coin is one that falls equally often on heads and tails. That ergodicity is a very weak form of randomness is shown by the example of a coin that falls according to a predictable sequence, e.g. heads followed by tails followed by heads, etc. In the same way, even regular orbits, which are highly predictable, are ergodic, since the time-averaged density of a regular orbit is constant on its torus (e.g. Binney & Tremaine 1987). *

A stronger, and physically more appealing, sort of random behavior is "mixing". A mixing system is one in which any small part of phase space will eventually spread itself uniformly over the entire accessible space, i.e. in which the coarse-grained density, evaluated at a *given* time, approaches a constant (e.g. Krylov 1979). Mixing is always associated with chaos, and regular orbits do not mix – any small patch on the torus simply translates, unchanged, around the torus. "Phase mixing" as understood by astronomers always involves sets of orbits on *different* tori, whose trajectories gradually move out of phase due to differences in period. Phase mixing is an intrinsically slow process and in fact has no well-defined time scale, since two regular trajectories that are sufficiently close will never pull apart.

---

* The term "ergodic" is also used to describe motion that visits every point on the energy surface – a definition that is too restrictive to be of much use for stellar dynamics.

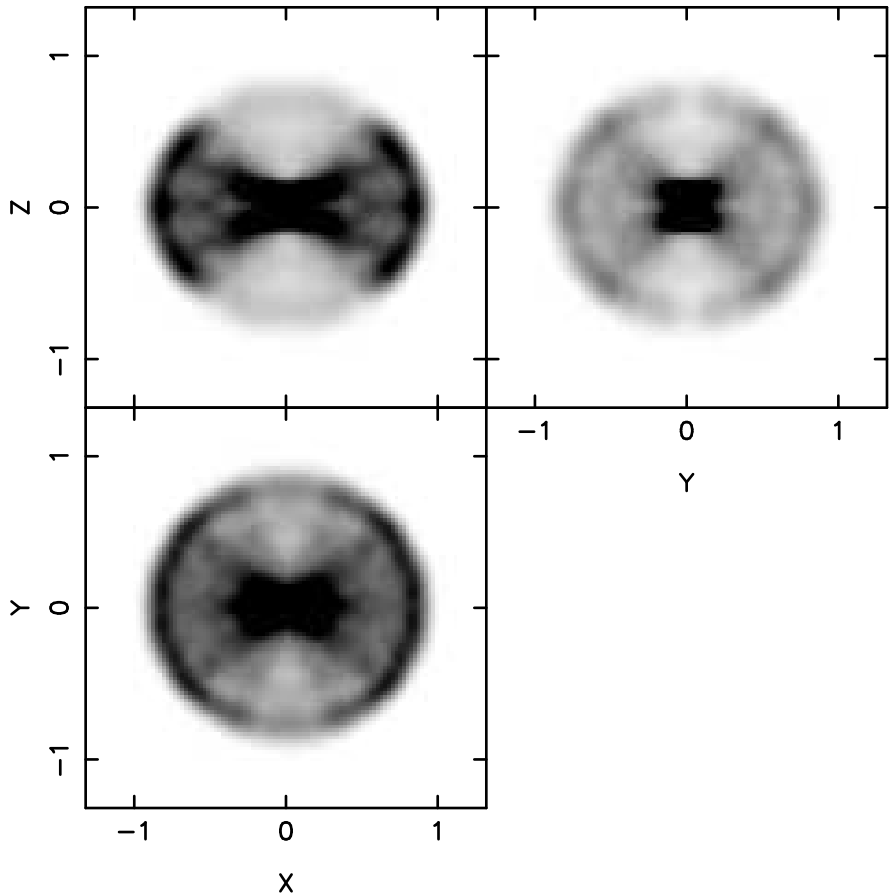

Fig. 3. Invariant density of an isoenergetic ensemble of $10^4$ stars in the chaotic phase space of a triaxial model with $m_0 = 10^{-3}$, $c/a = 0.5$ and $b/a = 0.79$. The $X$ [$Z$] axis is the long [short] axis of the triaxial figure. Plotted are the densities near each of the three principal planes.

Mixing implies the approach of an initially non-uniform phase space density toward a coarse-grained equilibrium. Fig. 3 shows the configuration-space density corresponding to such an equilibrium. It was computed by evolving a set of $10^4$ particles in the stochastic phase space of the model of Eq. (2), with $m_0 = 10^{-3}$, for 200 dynamical times; at this point the coarse-grained density defined by the ensemble had ceased to evolve significantly. The result – which might be called an "invariant density" (e. g. Kandrup & Mahon 1994) – looks remarkably like a regular box orbit: the density is elongated along the long ($X$) axis of the figure and is low around the short and intermediate axes. This is because stochastic orbits act, from turning point to turning point, very much like box orbits (Gerhard & Binney 1985), until they come close enough to the center to be perturbed onto another box-like trajectory, etc. Thus the invariant density corresponding to the full stochastic phase space at a given energy is similar to that of a superposition of box-like orbits (Merritt & Fridman 1996a).

The approach to an invariant, coarse-grained density via mixing can take place on a surprisingly short time scale, a point emphasized by Kandrup and coworkers (Kandrup & Mahon 1994; Mahon et al. 1995) who investigated chaotic motion in two-dimensional potentials. They estimated mixing time scales by coarse-graining the configuration space and computing the average deviation between the occupation numbers of an evolving ensemble, and the occupation numbers of the invariant density toward which the ensemble was evolving. The "distance" so defined decreased roughly exponentially with time, with a time constant of $\sim$ 20 dynamical times in a modified Plummer potential. Experiments in strongly stochastic triaxial potentials (Merritt & Valluri 1996) show that the mixing time scale is similar, of order $30-50$ dynamical times, or roughly ten Liapunov times — rather less than a galaxy lifetime. These results suggest that the stochastic orbits in the inner regions of triaxial galaxies ought to be "fully mixed." That is, the entire stochastic phase space at a given energy should be viewed as a single "orbit" with a well-defined shape and density distribution, like that of Fig. 3.

It is intriguing that one can define a mixing time scale for stochastic orbits but not for regular ones. It is nonetheless commonly assumed that the distribution of stars along regular orbits reaches an equilibrium state in just a few dynamical times after the formation of a galaxy. The mechanism that is usually invoked to produce this mixing – violent relaxation, i.e. the rapid mixing of phase space that takes place during the collapse and virialization of a galaxy (Lynden-Bell 1967) – is of course a chaotic process. But there is no obvious reason why violent relaxation should mix the regular parts of phase space while leaving the stochastic parts unmixed. Thus the most likely model for a galaxy might be one in which stochastic phase space is fully mixed from the start, in the same way that the regular parts of phase space are usually assumed to be. While this would be a difficult proposition to test, Habib et al. (1995) have shown how the addition of even a very small, time-dependent component to the potential can sometimes greatly enhance the mixing rate in stochastic systems. The much larger perturbations that are present during galaxy formation would presumably be even more effective.

## 5. Trapped Stochastic Orbits

In nearly integrable potentials, the stochastic orbits are strongly hampered in their motion by the surrounding invariant tori and can mimic regular orbits for many oscillations (e.g. Karney 1983). A certain number of these "trapped" stochastic orbits are expected to be present in any potential containing both regular and stochastic trajectories. Goodman & Schwarzschild (1981) investigated the motion in a weakly chaotic, triaxial potential with a large core and found that virtually all the stochastic orbits were trapped for $10^2$ dynamical times. They coined the term "semistochasticity" to describe this phenomenon.

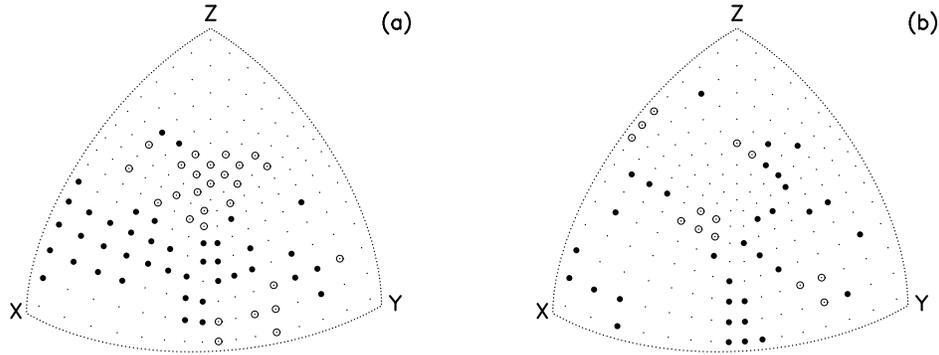

Fig. 4. Starting points of box-like orbits on one octant of the equipotential surface. Small dots are stochastic orbits; large dots are regular orbits; circles are trapped stochastic orbits. (a) Model of Fig. 1c ($m_0 = 10^{-3}$); (b) model of Fig. 2b ($m_0 = 10^{-1}; M_{BH} = 3 \times 10^{-3}$).

Trapped stochastic orbits are less important in the strongly chaotic triaxial models discussed here. Fig. 4 is a map of the regular and stochastic regions in initial condition space for the two triaxial potentials used to construct the histograms in Figs. 1c and 2b. Open circles represent starting points, on the equipotential surface, corresponding to trapped stochastic orbits; these were defined as orbits that have non-zero Liapunov numbers but which look essentially regular in configuration space for $10^3$ dynamical times or longer. (A roughly equivalent definition is: Orbits whose Liapunov numbers lie in the tail that extends leftward from the main stochastic peaks of Figs. 1c and 2b.) There are nearly as many of these trapped stochastic orbits as there are fully regular ones (although they comprise a small fraction of all stochastic orbits), which suggests that trapped orbits might play about as important a role as regular orbits in real galaxies. Similar numbers of trapped orbits are seen in triaxial models with weak cusps (Merritt & Fridman 1996a).

The mixing times quoted above were for ensembles of non-trapped, stochastic orbits. It seems certain that an ensemble of points in a trapped region of stochastic phase space would take a longer time to mix. The slow mixing of trapped stochastic orbits might introduce a long time scale into galaxy evolution.

## 6. Jeans's Theorem

Jeans's theorem states that the distribution function describing an equilibrium galaxy must depend only on the isolating integrals of motion. Jeans did not consider the case of a galactic potential containing both regular and chaotic orbits, but his theorem is easily generalized by introducing the concept of an "invariant density" defined above. Thus: an equilibrium galaxy is one that is representable as a superposition of time-invariant components. The latter include the uniformly-populated tori of regular phase space, but also the invariant densities that result from a uniform population of stochastic phase space at any energy.

As Fig. 3 illustrates, invariant densities in stochastic phase space can have considerable structure. This suggests that such components might be useful building blocks for real galaxies – though presumably not as useful as regular orbits, which are confined to smaller regions of phase-space and have a wider variety of shapes. Furthermore, if the relatively short mixing times described above are the rule in realistic triaxial potentials, then nature would be forced to incorporate the stochastic orbits via invariant densities, rather than via some more general, non-uniform population of stochastic phase space.

A rather different point of view was presented in a provocative paper by Binney (1982), who proposed that Jeans's theorem does not apply to systems containing stochastic orbits. Binney's idea was that mixing implies ergodicity, and only regular orbits can rigorously be shown to be ergodic. But regular orbits, while ergodic, do not mix; and it is mixing, not ergodicity, that is relevant for the approach to equilibrium. Thus one is almost tempted to turn Binney's argument on its head: stochastic orbits are "good" in the sense that they have a built-in mechanism that guarantees mixing; regular orbits are "bad" in the sense that their motion is quasi-periodic and hence non-mixing (at least in a fixed potential). In any case, neither ergodicity nor mixing appear to be essential properties of orbits in order for Jeans's theorem to be valid.

## 7. Self-Consistent Models

Before the publication of Schwarzschild's (1979, 1982) self-consistent models, triaxiality was generally considered to be an unlikely phenomenon because orbits in non-axisymmetric potentials were assumed to lack non-classical integrals of motion, i.e. to be chaotic. Schwarzschild showed that the motion in a triaxial potential with a large core is essentially fully regular, and that the four main families of regular orbits provided sufficient freedom to construct self-consistent equilibria. The work summarized above suggests roughly the opposite conclusion, namely that the motion in realistic, coreless triaxial potentials is largely chaotic. Does this mean that triaxial galaxies can not exist?

Schwarzschild (1993) himself first investigated this question by constructing self-consistent, scale-free ($\rho \propto r^{-2}$) triaxial models with six different choices of axis ratios. Schwarzschild's scale-free models were designed to represent the outer parts of galactic halos, and so he integrated individual orbits for only 55 dynamical times, roughly the age of the universe at $\sim 10$ kpc from the center of a galaxy. Many of his self-consistent solutions required the inclusion of stochastic orbits; however, since he allowed different stochastic orbits of the same energy to have different occupation numbers, the stochastic phase space was not uniformly populated and hence his models were not precisely stationary. Schwarzschild in fact showed that this non-uniform population of stochastic phase space implied that his models would evolve in shape in another 50 dynamical times.

Merritt & Fridman (1996a) constructed non-scale-free triaxial models with Dehnen's density law,

$$\rho(m) = \rho_0 m^{-\gamma}(1 + m)^{-(4-\gamma)} \quad (3)$$

both for $\gamma = 1$ ("weak cusp") and $\gamma = 2$ ("strong cusp"). Both models had $c/a = 0.5$ and $b/a = 0.79$. For neither model could self-consistency be achieved using only the regular orbits. Solutions constructed in the same way as Schwarzschild's (1993) – i.e. with arbitrary occupation numbers assigned to "different" stochastic orbits of the same energy – existed for both mass models, but real galaxies constructed in this way would evolve quickly near their centers due to mixing of the stochastic orbits. Solutions that were "fully mixed" near the center could be found for the weak-cusp model but not for the strong-cusp model. Thus, fully stationary phase-space populations do not always exist for triaxial mass models with strong central concentrations – real galaxies with the same density profiles would have to be weakly triaxial, or axisymmetric.

Although no one has yet attempted the construction of self-consistent triaxial models containing central black holes, Figures 2 and 4 suggest that chaos will play at least as strong a role in such models as it does in models with a strong cusp.

Given that stationary triaxial models are somewhat harder to construct than earlier believed, it is interesting to ask whether the *observational* evidence for triaxiality is compelling. The answer is probably "no", at least for the majority of elliptical galaxies (e.g. Gerhard 1994). The case is strongest for high-luminosity ellipticals, which have shallow cusps and low densities, hence long dynamical times – factors which would tend to reduce the mixing rates of stochastic orbits and thus to favor triaxiality. Low-luminosity ellipticals have steeper cusps and shorter dynamical times; interestingly, the kinematics of these galaxies have long been known to be crudely consistent with axial symmetry (Davies *et al.* 1983).

## 8. Acknowledgements

M. Valluri aided in the preparation of the figures. I thank her, H. Kandrup and G. Quinlan for useful discussions.

## 9. Discussion

*G. Contopoulos:* You have considered only nonrotating potentials. But if you include rotation, the classification of orbits is different. E.g. there are no box orbits, but close to the center there are elliptic orbits with loops. In our models of barred galaxies most orbits near the center are regular, while most chaotic orbits are near corotation.

*D. Merritt:* It is important to investigate whether slow figure rotation can strongly affect the degree of chaos in triaxial models. The fact that the axial orbits become loops when the figure rotates does not mean that the stochasticity will vanish, since an orbit need not pass exactly through the center in order to be stochastic.

# References

Binney, J. J.: 1982, *Mon. Not. R. Astron. Soc.* **201**, 15
Binney, J. and Tremaine, S.: 1987, *Galactic Dynamics*, Princeton University Press: Princeton, 171
Davies, R. L., Efstathiou, G., Fall, S. M., Illingworth, G. and Schechter, P. L.: 1983, *Astrophys. J.* **266**, 41
de Zeeuw, P. T.: 1985, *Mon. Not. R. Astron. Soc.* **216**, 273
de Zeeuw, P. T. and Lynden-Bell, D.: 1985, *Mon. Not. R. Astron. Soc.* **215**, 713
Ferrarese, L., van den Bosch, F. C., Ford, H. C., Jaffe, W. and O'Connell, R. W.: 1994, *Astron. J.* **108**, 1598
Ford, H. C., Harms, R. F., Tsvetanov, Z. I., Hartig, G. F., Dressel, L. L., Kriss, G. A., Bohlin, R. C., Davidsen, A. F., Margon, b. and Kochhar, A. K.: 1994, *Astrophys. J.* **435**, L27
Gebhardt, K.: 1995, private communication
Gerhard, O.: 1994, 'Elliptical Galaxies' in G. Contopoulos & A. Spyrou, eds., *6th European EADN Summer School: Galactic Dynamics and N-Body Simulations*, in press
Gerhard, O. and Binney, J. J.: 1985, *Mon. Not. R. Astron. Soc.* **216**, 467
Goodman, J. and Schwarzschild, M.: 1981, *Astrophys. J.* **245**, 1087
Habib, S., Kandrup, H. E. and Mahon, M. E.: 1995, *Phys. Rev. E*, in press
Kandrup, H. E. and Mahon, M. E.: 1994, *Phys. Rev. E* **49**, 3735
Karney, C. F. F.: 1983, *Physica* **8D**, 360
Kormendy, J., Dressler, A., Byun, Y.-I., Faber, S. M., Grillmair, C., Lauer, T. R., Richstone, D. and Tremaine, S.: 1995, 'An HST Survey of Early-Type Galaxies' in G. Meylan & P. Prugniel, eds., *ESO/OHP Workshop on Dwarf Galaxies*, ESO: Garching, 147
Krylov, N. S.: 1979, *Works on the Foundations of Statistical Physics*, Princeton University Princeton: Princeton
Kuzmin, G. G.: 1956, *Astr. Zh.* **33**, 27
Kuzmin, G. G.: 1973, 'Quadratic Integrals of Motion and Stellar Orbits in the Absence of Axial Symmetry of the Potential' in T. B. Omarov, eds., *Dynamics of Galaxies and Clusters*, Akad. Nauk. Kaz. SSR: Alma Ata, 71
Lauer, T., Ajhar, E. A., Byun, Y.-L., Dressler, A., Faber, S. M., Grillmair, C., Kormendy, J., Richstone, D. & Tremaine, S.: 1995, preprint
Lynden-Bell, D.: 1967, *Mon. Not. R. Astron. Soc.* **136**, 101
Mahon, M. E., Abernathy, R. A., Bradley, B. O. and Kandrup, H. E.: 1995, *Mon. Not. R. Astron. Soc.* **275**, 443
Merritt, D. and Fridman, T.: 1996a, *Astrophys. J.* **456**, in press
Merritt, D. and Fridman, T.: 1996b, 'Equilibrium and Stability of Elliptical Galaxies' in A. Buzzoni, A. Renzini & A. Serrano, eds., *Fresh Views of Elliptical Galaxies*, in press
Merritt, D. and Valluri, M.: 1996, in preparation
Miyoshi, M., Moran, J. Herrnstein, J., Greenhill, L., Nakai, N., Diamond, P. and Inoue, M.: 1995, *Nature* **373**, 127
Møller, P. Stiavelli, M. and Zeilinger, W. W.: 1995, *Mon. Not. R. Astron. Soc.* **276**, 979
Schwarzschild, M.: 1979, *Astrophys. J.* **232**, 236
Schwarzschild, M.: 1982, *Astrophys. J.* **263**, 599
Schwarzschild, M.: 1993, *Astrophys. J.* **409**, 563
Statler, T.: 1987, *Astrophys. J.* **321**, 113
Zaslavsky, G. M.: 1985, *Chaos in Dynamic Systems*, Harwood: London, 28